\documentclass[preprint2]{aastex631}

\usepackage{ae,aecompl}
\usepackage{subfigure}
\usepackage{graphicx}	
\usepackage{amsmath}	
\usepackage{amssymb}	
\usepackage{color,xcolor}
\usepackage{float}

\def   \aj {{\rm {AJ}}}
\def   \araa {{\rm {ARA\&A}}}
\def   \apj {{\rm {ApJ}}}

\def   \apjs {{\rm {ApJS}}}
\def   \apjl {{\rm {ApJL}}}

\def   \aap {{\rm {A\&A}}}

\def   \mnras {{\rm {MNRAS}}}
\def   \pasp {{\rm {PASP}}}

\def\pc{{\rm pc}}

\def\msun{M_\odot}

\def\kms{{\rm km\,s^{-1}}}

\def\yr{{\rm yr^{-1}}}
\def\sc{{\rm cm^{-2}}}
\def\cm2{{\rm cm^{-2}}}
\def\sqc{{\rm cm^{-2}}}

\def\micron{{\rm \mu m}}
\def\htwo{{\rm H_2}}

\def\hcn{{\rm HCN}}

\def\hctn{{\rm HC_3N}}

\def\hcop{{\rm HCO^+}}
\def\chtcn{{\rm CH_3CN}}
\def\htcop{{\rm H^{13}CO^+}}
\def\chtoh{{\rm CH_3OH}}
\def\htcn{{\rm H^{13}CN}}

\def\cch{{\rm CCH}}

\def\mjbm{{\rm mJy\, beam^{-1}}}

\graphicspath{{./}{figures/}}

\received{January xx, 2022}
\revised{January xx, 2022}
\accepted{\today}
\submitjournal{ApJ}

\shorttitle{Massive Escaping YSO}
\shortauthors{Ren et al.}

\begin{document}
\title{A High-Mass Young Star-forming Core Escaping from Its Parental Filament}


\correspondingauthor{Zhiyuan Ren}
\email{renzy@nao.cas.cn}

\author[0000-0003-4659-1742]{Zhiyuan Ren}
\affiliation{National Astronomical Observatories, Chinese Academy of Sciences, Datun Rd. A20, Beijing, People's Republic of China}
\affiliation{CAS Key Laboratory of FAST, NAOC, Chinese Academy of Sciences, Beijing, China; University of Chinese Academy of Sciences, Beijing, China}

\author[0000-0002-5435-925X]{Xi Chen}
\affiliation{Center for Astrophysics, Guangzhou University, Guangzhou 510006, People's Republic of China}
\affiliation{Shanghai Astronomical Observatory, Chinese Academy of Sciences, 80 Nandan Road, Shanghai 200030, People's Republic of China}

\author[0000-0002-5286-2564]{Tie Liu}
\affiliation{Shanghai Astronomical Observatory, Chinese Academy of Sciences, 80 Nandan Road, Shanghai 200030, People's Republic of China}

\author{Emma Mannfors}
\affiliation{Department of Physics, P.O. box 64, FI- 00014, University of Helsinki, Finland}

\author{Leonardo Bronfman}
\affiliation{Departamento de Astronom\'{i}a, Universidad de Chile, Las Condes, Santiago, Chile}

\author[0000-0001-5950-1932]{Fengwei Xu}
\affiliation{Kavli Institute for Astronomy and Astrophysics, Peking University, 5 Yiheyuan Road, Haidian District, Beijing 100871, People’s Republic of China}

\author[0000-0002-4707-8409]{Siyi Feng}
\affiliation{Department of Astronomy, Xiamen University, Zengcuo'an West Road, Xiamen, 361005, People's Republic of China}

\author[0000-0003-3343-9645]{Hongli Liu}
\affiliation{Department of Astronomy, Yunnan University, Kunming, 650091, PR China}

\author[0000-0002-5927-2049]{Fanyi Meng}
\affiliation{University of Chinese Academy of Sciences, Beijing 100049, China}

\author[0000-0003-2300-8200]{Amelia M.\ Stutz}
\affiliation{Departamento de Astronom\'{i}a, Universidad de Concepci\'{o}n,Casilla 160-C, Concepci\'{o}n, Chile}

\author[0000-0003-1275-5251]{Shanghuo Li}
\affiliation{Max Planck Institute for Astronomy, K\"{o}nigstuhl 17, D-69117 Heidelberg, Germany}

\author[0000-0002-3179-6334]{Chang Won Lee}
\affiliation{Korea Astronomy and Space Science Institute, 776 Daedeokdae-ro, Yuseong-gu, Daejeon 34055, Republic of Korea}
\affiliation{University of Science and Technology, Korea (UST), 217 Gajeong-ro, Yuseong-gu, Daejeon 34113, Republic of Korea}

\author[0000-0002-7237-3856]{Ke Wang}
\affiliation{Kavli Institute for Astronomy and Astrophysics, Peking University, 5 Yiheyuan Road, Haidian District, Beijing 100871, People’s Republic of China}

\author{Jianwen Zhou}
\affiliation{Kavli Institute for Astronomy and Astrophysics, Peking University, 5 Yiheyuan Road, Haidian District, Beijing 100871, People’s Republic of China}

\author[0000-0003-3010-7661]{Di Li}
\affiliation{National Astronomical Observatories, Chinese Academy of Sciences, Datun Rd. A20, Beijing, People's Republic of China}
\affiliation{CAS Key Laboratory of FAST, NAOC, Chinese Academy of Sciences, Beijing, China; University of Chinese Academy of Sciences, Beijing, China}
\affiliation{NAOC-UKZN Computational Astrophysics Centre (NUCAC), University of KwaZulu-Natal, Durban 4000, South Africa}

\author{Chen Wang}
\affiliation{National Astronomical Observatories, Chinese Academy of Sciences, Datun Rd. A20, Beijing, People's Republic of China}

\author[0000-0003-4761-6139]{Chakali Eswaraiah}
\affiliation{Indian Institute of Science Education and Research (IISER) Tirupati, Rami Reddy Nagar, Karakambadi Road, Mangalam (P.O.), Tirupati 517 507, India}

\author[0000-0001-5917-5751]{Anandmayee Tej}
\affiliation{Indian Institute of Space Science and Technology, Thiruvananthapuram, Kerala 695 547, India}

\author[0000-0002-9703-3110]{Long-Fei Chen}
\affiliation{Research Center for Intelligent Computing Platforms, Zhejiang Laboratory, Hangzhou 311100, China}

\author{Hui Shi}
\affiliation{National Astronomical Observatories, Chinese Academy of Sciences, Datun Rd. A20, Beijing, People's Republic of China}



\begin{abstract}
We studied the unique kinematic properties in massive filament G352.63-1.07 at $10^3$-AU spatial scale with the dense molecular tracers observed with the Atacama Large Millimeter/submillimeter Array (ALMA). We find the central massive core M1 ($12 \msun$) being separated from the surrounding filament with a velocity difference of $v- \overline{v}_{\rm sys}=-2\,\kms$ and a transverse separation within 3 arcsec. Meanwhile, as shown in multiple dense-gas tracers, M1 has a spatial extension closely aligned with the main filament and is connected to the filament towards its both ends. M1 thus represents a very beginning state for a massive young star-forming core escaping from the parental filament, within a time scale of $\sim 4000$ years. Based on its kinetic energy ($3.5\times10^{44}$ erg), the core escape is unlikely solely due to the original filament motion or magnetic field, but requires more energetic events such as a rapid intense anisotropic collapse. The released energy also seems to noticeably increase the environmental turbulence. This may help the filament to become stabilized again. 
\end{abstract}



\section{Introduction} \label{sec:intro}
Star-forming clouds can cause several prominent dynamical consequences. One is the escaping and spreading of young stars towards the surrounding space, which has caught increasing attention during the recent years \citep[e.g.][]{tobin09,kraus17,herczeg19,gaudin19,swiggum21}. The stars tend to be gradually drifting over the entire cloud, forming a similar but more extended spatial distribution \citep{grobschedl18,jerabkova19,kuhn20,ward20,gupta22}. This provides a unique proxy to inspect the cloud morphology and mass distribution. Furthermore, based on the recent Gaia-based measurement of the trigonometric parallax and stellar proper motion, one can explore the spatial-velocity structure with an unprecedented accuracy and a historical perspective \citep[e.g.][]{szegedi19,roccatagliata20,galli20,krolikowski21,kounkel22,tu22,ha22,dharmawardena22,duan23}. These works have largely improved our understanding of the dynamical condition and evolutionary trend of the star-forming clouds.

As one critical but incomplete aspect of this study, the initial condition of the escaping young stars is still not widely inspected. With the exception of some extremely high-velocity stars (10 to $10^2\,\kms$) ejected from n-body ($n\geq 3$) interactions \citep{ducourant17,fernandes19,bally20,ortiz21}, the stellar motions could also be affected by their dense-gas structures. Young stars could directly inherit the turbulent field of the gas structure \citep{ha21,krolikowski21,quintana22,gupta22}, and obtain further acceleration due to the gravitational instability or other interactions among the parental filaments or clumps \citep{stutz16,getman19,zamora19,kim19,sharma20,alvarez21}. \citet{stutz16} proposed one particular scenario where the filament in Orion A could have an oscillation that drives the young stars towards transverse directions, eventually forming a broadened stellar distribution in parallel with the filament. But in all these studies, the young stars are either totally separated from the filaments or have dissipated the surrounding gas \citep[e.g.][]{jerabkova19,gupta22}. So we still need to investigate the initial dynamical condition when the YSOs are leaving the parental structures.

G352.63-1.07 is a massive young star forming region at a precisely measured distance of $D=690$ pc \citep[][Chen21 hereafter]{chen21}. It contains several massive cores aligned on a dense filament. As shown in Chen21, the filament has actively ongoing dynamical features, including prominent outflows and mass transfer flows, which induced bright shock emissions. However, due to the limited molecular tracers and sensitivities, the gas motions among the cores and filaments are still undetermined. In this work, we explored the gas motions using optically thin lines, which demonstrate the unique tendency of the escaping massive core M1. The observed data is described in Section \ref{sec:obs}. The global dense-gas and velocity distribution is reported in Section \ref{sec:results}. The dynamical origin and the time-energy properties of the gas motion are discussed in Section \ref{sec:discussion}. A summary is given in Section \ref{sec:summary}.     

\section{Observation} \label{sec:obs}
G352.63-1.07 (G352 hereafter) is observed with the Atacama Large Millimeter/submillimeter Array (ALMA) in the ATOMS survey \citep{liu20,liu22}. The observation was performed with both Compact 7-m Array (ACA) and the 12-m array (C43-2 or C43-3 configurations) in Band 3. The restored data cube has an average beam size of $\theta_{\rm maj}\times\theta_{\rm min}=2''.5\times2''.1\ (PA=-87^\circ)$. The antenna baselines can cover the extended structures with a scale up to $\sim100''$. The channel width and noise level varies with different spectral windows. The $\htcop$, $\cch$, and $\htcn$ (1-0) lines have $\Delta v_{\rm chan}=0.2\,\kms$, $\sigma_{\rm rms}=5\,\mjbm$ (0.15 K). The $\hcop$ (1-0) has $\Delta v_{\rm chan}=0.1\,\kms$, $\sigma_{\rm rms}=9\,\mjbm$ (0.25 K). The $\chtoh$ ($2_{11}-1_{10}$) lines have $v_{\rm chan}=1.4\,\kms$, $\sigma_{\rm rms}=2\,\mjbm$ (0.08 K). The more detailed observing conditions are presented in \citet{liu20}.     

\section{Results} \label{sec:results}
\subsection{Filament and cores}  \label{sec:results_cores}
Figure \ref{fig:spec_img}a shows the 3 mm continuum emission and the integrated $\htcop$ (1-0) emission (moment 0). Compared to the chemically fresh molecules $\cch$ and $\hctn$ (Chen21, Figure 3 therein), the $\htcop$ emission is more concentrated towards the inner region of the filament structures. It shows an S-shaped compact filament. The six major cores (Chen 21) are clearly resolved therein. The line intensity decreases towards M4 and M5, which could be due to the gas dissipation caused by their outflows (Appendix \ref{app:lines}). 

Figure \ref{fig:spec_img}b shows the $\htcop$ emission overlaid on the Spitzer/IRAC three-band color image \citep[GLIMPSE survey,][]{Benjamin03, Churchwell09}. Among the dense cores, only M1 has a compact IR source. M2 to M5 are all absent of IR point sources. The other two IR sources (M3b and M6) are located on the western side of the filament. The IRAC color magnitudes for M1 are [3.6]-[4.5]=2.1 and [5.8]-[8.0]=1.0, which is similar to the color of Class-0 YSOs \citep{megeath12}. M1 is also detected in 6.7 GHz methanol maser but undetected in 6-8 GHz radio continuum emission above the noise level of 3-8 $\mjbm$ \citep{walsh98}. This indicates M1 to have a deeply embedded young massive star. It has not yet caused significant ionization to the surrounding medium. Figure \ref{fig:spec_img}b also shows the $\chtoh$ (2-1) emission. It has a compact morphology concentrated at M1. 

The molecular spectra at the three inner cores are shown in Figure \ref{fig:spec_img}c. The $\htcop$ (1-0) exhibits a noticeable double-peak profile at M1. The two components have a separation of $v_{\rm blue}-v_{\rm red} = -1.8 \,\kms$. M2 and M3 both have a single-velocity component within $v_{\rm lsr} = \pm 0.3\,\kms$, which are close to the redshift component at M1. The $\hcop$ and $\chtoh$ line profiles are less resolved owing to high optical depth and low spectral resolution, respectively. But we can still see these two lines inclined to the blueshift side. We also examined the other two dense-gas tracers CCH and $\htcn$ (1-0) (Figure \ref{fig:pv_3mol}). They have similar velocity components to the $\htcop$ (1-0). The molecular lines all indicates prominent blueshift component at M1.   

\subsection{Physical parameters and gravitational instability} \label{sec:pars} 
The physical parameters of the cores are presented in Table \ref{tab:cores}. The details of the calculation are described in Appendix \ref{app:parameters}. All the cores have comparable spatial sizes are nearly uniformly distributed along the filament, with an average spatial interval of $\Delta s_{\rm cores}=(10\pm 2)''$, or $(7\pm1.5)\times 10^3$ AU. 

From the observed parameters, we estimated the stability of the cores from the critical mass \citep{bertoldi92,mckee07,li13}:
\begin{equation}\label{equ:m_crit}
\begin{aligned}
M_{\rm crit} & = M_{\rm BE} + M_{\rm \Phi} \\
\quad        & = \frac{5R_{\rm eff}}{G}\left(\sigma_{\rm tot}^2 + \frac{v_A^2}{6} \right),
\end{aligned}
\end{equation}

wherein $\sigma_{\rm tot}$ is the effective velocity dispersion (see Appendix \ref{app:parameters}), $M_{\rm BE}$ and $M_{\rm \Phi}$ represent the upper-limit masses to be supported by the turbulence and magnetic pressures, respectively. $M_{\rm \Phi}$ depends on magnetic field strength $B$ and mass density $\rho_0=\mu m_{\rm H} n_0$, which are included in Alfv\'{e}n velocity of $v_A=B/\sqrt{4\pi \rho_0}$. For the $B$ value, we referred to the B-field measurement in other similar cold dense filaments \citep[e.g.][]{liu18b,ching18}. They are measured to have a bulk distribution of $B=0.5$ to 1.0 mG. Adopting this range, we can derive $M_{\rm \Phi}=0.6$ to $1.2\msun$, which only has a minor contribution to $M_{\rm crit}$. As shown in Table \ref{tab:cores}, all the cores have $M_{\rm crit} > M_{\rm core}$, which indicates a subcritical state if the turbulence can support them against the self-gravity.

The filament stability can be estimated from the critical line-mass density \citep{ostriker64,arzoumanian13}
\begin{equation}\label{equ:eta_fil}
\left(\frac{M}{l}\right)_{\rm crit} = \frac{2 \sigma_{\rm tot}^2}{G}.
\end{equation} 
From the average line width of $\Delta v_{\rm fila}(\htcop)=2.0\,\kms$, we can derive $\sigma_{\rm tot}=0.9\,\kms$ and $(M/l)_{\rm crit}\simeq 390\,\msun\,{\pc}$. In comparison, the observed total mass ($48\,\msun$) and length ($45''$) yield $(M/l)_{\rm obs}\simeq 300\,\msun\,\pc^{-1}$. Considering the projection effect, the actual value could be even smaller. Therefore the property of $(M/l)_{\rm obs}<(M/l)_{\rm crit}$ suggests a subcritical state also for the entire filament. The turbulent condition of the filament and cores should have a close interplay with other dynamical features including the core collapse and escaping motion, as discussed bellow.

\section{Discussion} \label{sec:discussion}
\subsection{Resolving the velocity components} \label{sec:origin}
As shown above, the filament has an overall smooth and compact morphology. The transfer flows (Chen21) and outflows seem to only have a limited influence to the main filament. As seen in optically thin lines, the blueshift motion of M1 is still the most conspicuous feature over the entire structure. Its dynamical origin should be further inspected.

Figure \ref{fig:m1_blue}a shows the $\htcop$ (1-0) emission in four different velocity channels. The ALMA 3 mm and SMA 1 mm (Chen21) dust continuum emissions are also plotted in sub-panels for comparison. Figure \ref{fig:m1_blue}b presents the position-velocity (PV) plot of the $\htcop$ along the filament. It shows that the blue component is mainly in the velocity range from -4 to -1 $\kms$ and has a spatial extension from offset=$-16''$ to $+6''$. In order inspect the gas motion, we plot the emission region in two velocity intervals, denoted as low- and high-velocity components, or LVC and HVC, respectively. we plot their emission regions in Figure \ref{fig:m1_blue}a. One can see that the LVC is peaked at M1 and has a weak elongation towards M2, while the HVC is more confined around M1. Their morphologies are both largely different from the outflow lobes in Figure \ref{fig:outflow_spec}a. They should both trace the dense-core motion instead of outflow. 
 
The filament component is mainly in the velocity range of (-1.5,+1.5) $\kms$. Its emission region (false-color image in Figure \ref{fig:m1_blue}a) almost traces the entire gas structure, and shows a noticeable gap at M1 center. The redshift wing component (1.5 to 2.2 $\kms$, yellow contours) includes two separated patches, with one extending from M4 and M5 and another around M3. They could trace the denser gas affected by the outflow or transfer flow.  Around M1, there are no evident redshift feature at $v>1\ \kms$. The dust continuum emissions (Figure \ref{fig:m1_blue}a, two right panels) show an overall similar spatial extent with the $\htcop$ filament, but have a compact intensity peak rightly at M1, which is $\sim 3$ times more intense than the remaining filament.

The spatial correlation of the velocity components can also be inspected from their intensity profiles as shown in Figure \ref{fig:m1_blue}b (lower panel). The dust continuum profiles (dashed and dotted lines) rightly follow the blueshift component (blue solid line). They all have a shoulder-like decreasing trend from M1 to M2. This coherency suggests that the dense core M1 should be mainly associated with the blueshift gas. 

The intensity gap on the filament is more clearly seen in the intensity profile (red line in Figure \ref{fig:m1_blue}b, lower panel). It rightly coincides with the M1 center of both the dust continuum and the blueshift $\htcop$ profiles. For the $\htcop$ emission, if considering the total intensity profile (black solid line), the blueshift component would nicely fill the gap, making the entire profile much more flattened. 

The CCH emission regions of the two components are shown in Figure \ref{fig:pv_3mol}a. The PV diagrams of CCH and $\htcn$ are shown in Figure \ref{fig:pv_3mol}b. The lower intensity around M1 along the filament can also be seen in CCH (Figure \ref{fig:pv_3mol}a). Although its blue component is more extended than that of $\htcop$, it is still mainly concentrated at the central cavity of the filament. The $\htcn$ emission is more concentrated at M1, so the central intensity decline is not evident. But from its PV diagram, one can still see a comparable velocity separation of $v-v_{\rm lsr}=-2\ \kms$ between the blueshift component and the main filament. And it is also closely overlapped with the $\htcop$ and CCH spectra for the blueshift peak (Figure \ref{fig:outflow_spec}c). The three molecular lines thus consistently suggest the blueshift component to be the dominant one at M1. In other words, the dense core is having a bulk motion relative to the rest part of the filament.    

Based on their spatial and velocity features, M1 and the main filament could have two possible configurations. M1 can be either leaving the filament on the front side or moving towards it from behind. From the PV plot (Figure \ref{fig:m1_blue}b), one can see the blue component connected to the main filament both towards north and south, around offset=$+5''$ and $-15''$, respectively. The PV diagrams of the CCH and $\htcn$ (1-0) lines (Figure \ref{fig:pv_3mol}, right column) show the similar connection feature between M1 and the main filament in the two directions. M1 should thus be originally formed in the filament, and have obtained the blueshift motion only during the recent time.  

\subsection{Driving force of the blue component} \label{sec:origin}
The multiple velocity components in G352 are comparable to other filaments with prominent kinematical features. As shown in previous studies, undisturbed linear filaments tend to have moderate fluctuation of $\vert v-v_{\rm sys} \vert \leq 1.0\,\kms$ around their internal cores \citep[e.g.][]{punanova18,bhadari20}. More complicated fiber-composed filaments also have similar velocity variation within $1\,\kms$ \citep{hacar17,Clarke18}. Stronger kinematical features of several $\kms$ are seen in filaments with collapse \citep{henshaw14,chen19a,ren21,li22a,cao22} or interactions \citep{shimajiri19,anathpindika21}. The velocity fluctuations in G352 resembles those most active ones. The blueshift motion ($-2\ \kms$) and its spatial scale ($5''$ or 3500 AU) lead to a gradient of $\sim 250\,\kms \pc^{-1}$, which resembles those most intensely collapsing filaments \citep[e.g.][]{peretto13,Montillaud19,hu21,Beuther21,cao22}.

In collapsing or fiber-composed filaments, the global velocity field is often entangled with the individual core motions. It is therefore uncertain if the cores are co-moving with the filament or already separated. In comparison, G352 presents an example of clearly separated velocity components. In fact, the filament is also inclined to the blueshift side around M1 towards -0.5 $\kms$ (Figure \ref{fig:m1_blue}b). The filament could thus have a co-moving tendency with M1. In particular, as shown in Figure \ref{fig:m1_blue}a, the HVC ($v=(-4,-3)\ \kms$) is narrowly confined between the northern and southern segments of the entire filament. This increases the evidence that the core and filament motions are closely related.

The mass transfer flows (Chen21) provide a viable mechanism to initiate the filament collapse. Since the flows are observed from one-sided molecular line wings instead of strong infall signatures, they would provide a moderate mass accumulation to the inner filament (between M2 and M3). If the mass assembly once exceed the threshold of $(M/l)_{\rm crit}$ or $M_{\rm crit}$, it would possibly induce a major collapse. One can see that the transfer flows exhibit no evident mass assembly at their arrival points on the inner filament (Figure \ref{fig:m1_blue}). This also indicates that the flows should have a further propagation towards M1. 

\subsection{Energy scales of the gas components} \label{sec:energy}
The filament collapse and core escape can also be examined from their energy scales. From the core mass and its escaping velocity, the kinetic energy of M1 is estimated to be $E_{\rm k}= m_{\rm core}(v_{\rm blue}-v_{\rm red})^2/2\simeq3.5\times 10^{44}$ erg. This would represent a lower limit due to the projection effect. In comparison, the turbulent energy of the entire filament is $E_{\rm turb}=m_{\rm tot}\sigma_{\rm nt}^2/2=(2.5\pm 0.5)\times 10^{44}$ erg. As for the magnetic energy, from the speculated $B$-range (Section \ref{sec:pars}), we can estimate $E_B=m_{\rm tot} v_A^2/2=2$ to $8\times10^{43}$ erg. $E_{\rm turb}$ and $E_B$ values characterize the entire filament. The fraction to reach M1 could be even lower thus cannot be solely responsible for the core motion.

The collapsing energy can be estimated as $\Delta E_{\rm p,coll} = G m_{\rm in}^2[(1/r_1)-(1/r_0)]$, which involves the initial and final radii ($r_0$ and $r_1$) of the collapsing mass $m_{\rm coll}$. Assuming that the major collapse is taking place between M2 and M3, we can adopt $r_0=\Delta s_{\rm cores}$. Together with the current radius and mass of M1, we derived $\Delta E_{\rm p,coll} \simeq 6\times 10^{44}$ erg, which could be sufficient to accelerate M1.

As seen in PV diagram (Figure \ref{fig:m1_blue}b), the inner region has a typical line width of $\Delta v\geq 1.8\kms$ for both the blueshift component and the main filament. It rapidly declines towards the outer part, reaching a trans-sonic level of $\Delta v\simeq 0.7\ \kms$ ($\sigma_{\rm nt}=0.4\,\kms$) towards offset=$\pm 20 ''$. The low-$\Delta v$ areas over the filament could represent the fraction less affected by the collapse. Adopting the average value ($0.9\ \kms$) in Equation \ref{equ:eta_fil}, we can derive $(M/l)_{\rm crit}\simeq 150\,\msun\,\pc^{-1}$, which is much smaller than $(M/l)_{\rm obs}$ and should reflect the supercritical condition before the major collapse. 

Figure \ref{fig:mom_3mol} shows the velocity distributions over the core and filament in all three lines. The low turbulence areas with $\Delta v$ down to $0.6\ \kms$ can be seen in all three species. The CCH and $\htcop$ emission regions both exhibit the low values down to $\Delta v\simeq 0.7\ \kms$, while the higher values mainly appear between M1 and M3. The $\htcn$ line is more concentrated on dense cores, which could be due to its chemical bias towards the protostellar stage. Along the filament, the $\htcn$ line width variation is mainly between 1.2 and 1.4 $\kms$, which is comparable to average values of CCH and $\htcop$ lines.    

\subsection{Time scales of the gas motion} \label{sec:picture}
Figure \ref{fig:ridge_dv}a shows the ridge lines of the main filament and the blueshift component (M1 and southern extension). The two components have a transverse offset of $\Delta s_{\rm esc}=0$ to 2.3 arcsec along the filament. We plot the $\Delta s_{\rm esc}$ and $\vert v_{\rm blue}-v_{\rm red}\vert$ profiles along the filament in Figure \ref{fig:ridge_dv}b. One can see that the two quantities have a coherent variation trend. At M1 and the second velocity peak near M2 (offset=$-10''$), the velocity and transverse offsets both reach a local maximum. 

The transverse separation as a function of time can be estimated as $\Delta s_{\rm esc}=t_{\rm esc}(v_{\rm blue}-v_{\rm red})\tan \theta $, wherein $\theta$ is the inclination angle between the gas motion and sightline. Adopting $\theta=(45\pm10)^\circ$, we can derive a timescale of $t_{\rm esc}\simeq (4\pm 2) \times 10^3$ years for the blueshift motion. It is notably close to the outflow age of $t_{\rm out}=3.6\times 10^3$ years (Appendix \ref{app:parameters}). The two values should together characterize the star-forming age in G352.

If the core escape started during the filament collapse, one would expect an overall collapsing time sufficiently longer than $t_{\rm esc}$. From the recent semi-analytical work \citep{Clarke15}, the filament could have a collapsing time of
\begin{equation}\label{equ:eta_fil}
t_{\rm coll} = (0.49 + 0.26 A_0) (G \rho_0)^{-1/2},
\end{equation} 
wherein $A_0$ is the aspect ratio of the filament. Assuming that the major collapse took place in the inner region between M2 and M3, from the filament length ($l \simeq 22''$) and width ($w \simeq 5''$), we can derive $t_{\rm coll}\simeq 6\times 10^4$ years. It could roughly represent the dense-core formation time. For M1, it then implies a mass accretion rate of $\dot{M}_{\rm acc} = M_{\rm core}/t_{\rm coll} \simeq 1.8\times 10^{-4}\,\msun\,\yr^{-1}$, which is comparable to the rate of the transfer flows (chen21). The collapse towards center could be dominated by the two colliding flows along the filament. The flows would rapidly increase the central density and develop an anisotropic instability. The highly compressed gas would then move towards the transverse directions, initiating the observed motion. A schematic view of the collapse and core acceleration is shown in Figure \ref{fig:ridge_dv}c.  

The collapsing time could be compared with the turbulence-increasing time of the filament. From the turbulence level of the inner filament ($\sigma_{\rm nt}\simeq0.8\,\kms$), we can derive a dynamical timescale of $t_{\rm dyn}=\Delta s_{\rm cores}/\sigma_{\rm nt}\simeq 8\times 10^4$ years, which is also similar to $t_{\rm coll}$. As shown in Figure \ref{fig:mom_3mol}, the line-width distributions of $\htcop$ and CCH are both steeply increased between M1 and M3 with a scale of $\Delta V \geq 1.0\,\kms$. This also provides evidence that the turbulence was increased by the recent dynamical evolution thus have not yet reached a uniform state over the filament.   

Although the filament collapse could have sufficient energy and time to initiate the core motion, other dynamical origins for the blue component are still not fully excluded. In particular, it could also be a fiber-like component. The fibers in massive filaments are observed on larger spatial scales ($>0.2$ pc) in a few recent studies \citep{shimajiri19,cao22}. Those fibers are only loosely aligned or intertwined along the filament axis. They have transverse separations up to several 0.1 pc, and do not have two-ends connections like in the case of M1. The fibers could be increasingly separated with the filament age. For G352, the separation would be $\Delta s_{\rm esc} \geq (v_{\rm blue}-v_{\rm red}) t_{\rm coll} \simeq 2.5\times 10^4$ AU ($36''$), which is unreasonably large compared to the observed spatial scales. To maintain the observed $\Delta s_{\rm esc}$, the fibers should have originally much smaller velocity difference ($\ll 2\,\kms$), and still need a recent collapse to provide the kinetic energy and reach the escaping velocity.    

\section{Summary} \label{sec:summary}
Based on the velocity distribution of $\htcop$ and other dense-gas tracers, we found that the most massive core M1 ($12\,\msun$) is connected to the main filament, but has a blueshift of $v_{\rm core} - \bar{v}_{\rm sys}\simeq-2\,\kms$. The core motion has a kinetic energy of $3.5\times 10^{44}$ erg and timescale of only $\sim 4000$ years. As a unique example, M1 reveals the very beginning of a massive YSO escaping from its birth site. It confirms the possibility that YSO can leave the filament along the transverse direction.

Among the available mechanisms, only the filament collapse could provide enough energy to initiate the core escape. A recent collapse is also needed to generate the massive star therein. It can be also responsible for the drastic velocity and turbulence variations over the filament. As the core motion is taking place, a central gap is formed rightly at its location on the remaining part of the filament.  

According to the energy and time comparisons, the process of core escape could have three major steps: (i) the filament had a relatively low turbulence, with moderate contraction and fragmentation into the dense cores; (ii) because of the transfer flows, the inner region became more massive and unstable, and eventually initiated a major collapse; (iii) M1 was strongly compressed by the collapsing gas to start the escaping motion, while the rest energy could be transferred to the filament to increase the turbulence. 

Based on the result in G352, one can examine other filaments with steep velocity gradients to see if similar core motions are taking place. A larger sample will help evaluate the universality of such collapse-induced YSO escape, and estimate its contribution to the YSO dispersion over molecular clouds. 

\clearpage

\begin{table*}[t]
\centering
\begin{minipage}{140mm}
\caption{The physical properties of the cores. }
\begin{tabular}{lcccccc}
\hline\hline
Parameters                                         &   M1             &   M2            &   M2b           &   M3               &  M4              &  M5      \\
\hline                                                                                                                                            
\multicolumn{7}{l}{\bf (Observed)}  \\                                                                                                 
$v_{\rm lsr}\ (\htcop)$ ($\kms$)                   &  -2.5            &   -0.8          &  -0.6           &  +0.1              &  +1.0             &  +1.3     \\ 
$T_{\rm ex}\ (\hcop)$ (K)                          &  37              &   22            &   22            &   22               &  22              &  20      \\ 
$T_{\rm b}\ (\htcop)$ (K)                          &  4.0             &   3.0           &   5.0           &   2.1              &  3.7             &  2.6     \\
$\Delta v$ ($\kms$)                                &  2.5             &   2.3           &   1.8           &   2.2              &  1.8             &  1.7     \\         
Radius (arcsec)$^a$                                &  5               &   4             &   6             &   5                &  4               &  4       \\       
\multicolumn{7}{l}{\bf (Derived)}   \\                                                                           
$N_{\rm tot}$ ($10^{23}\,\cm2$) $^b$               &  $2.6\pm0.3$     &  $2.2\pm0.2$    &  $2.3\pm0.3$    &   $2.4\pm0.2$      &  $2.2\pm0.2$     &  $2.2\pm0.2$ \\  
$\sigma_{\rm tot}$ ($\kms$)                        &  1.1             &  0.9            &  0.8            &   1.0              &  0.7             &  0.7      \\          
Mass ($\msun$)                                     &  $12\pm2$        &  $6\pm2$        &  $11\pm2$       &   $8\pm2$          &  $7\pm2$         &  $7\pm2$   \\         
$m_{\rm crit}$ ($\msun$)                           &  15              &  13             &  12             &   12               &  11              &  11 \\                
\hline
\end{tabular} \\
a.{ Average radius deconvolved with the beam size. }  \\
b.{ $N_{\rm tot}$ is derived from $\htcop$ intensity using Equation \ref{equ:n_tot}. For M4 and M5, the $\htcop$ emissions are noticeably dissipated by the outflows. We assumed them to have same $N_{\rm tot}$ with M3 based on the fact that these cores also have comparable 3 mm continuum emissions. } \\
\label{tab:cores}
\end{minipage}
\end{table*}

\begin{figure*}[t]
    \centering
    \includegraphics[angle=0, width=0.95\textwidth]{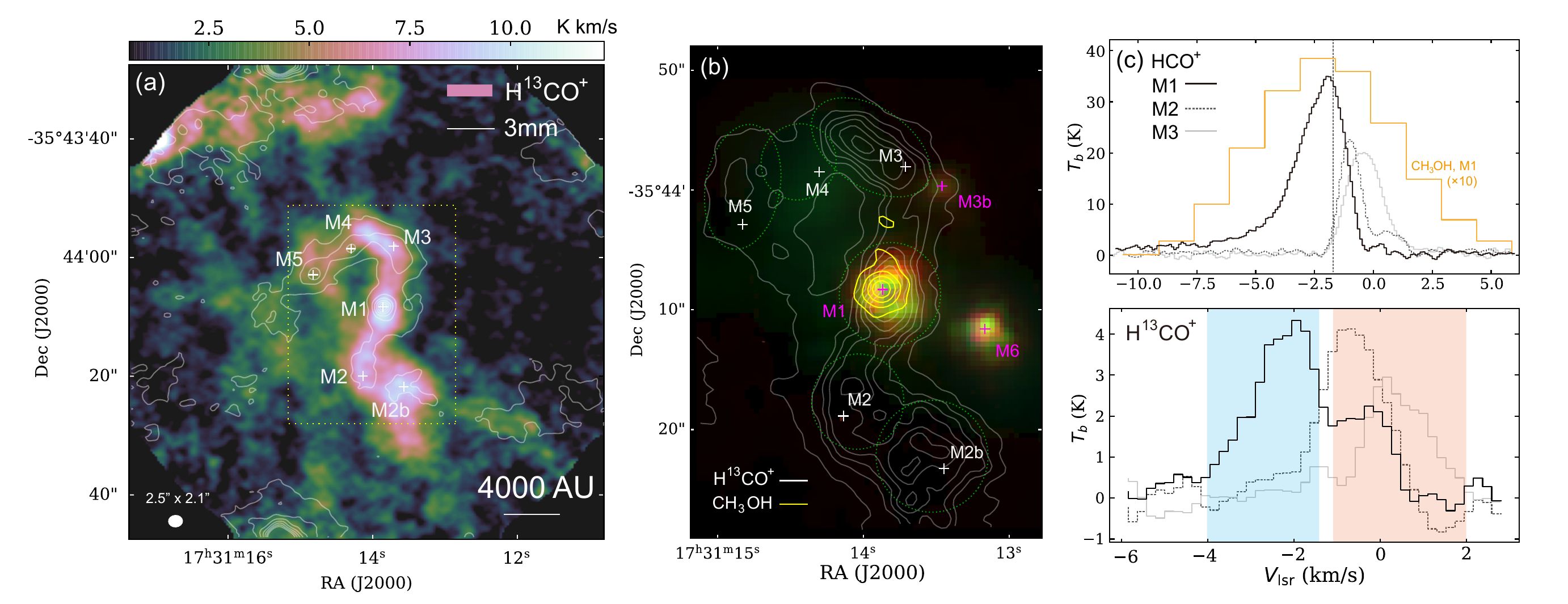} \\
    \caption{\small {\bf (a)} The velocity-integrated intensity of the $\htcop$ (1-0) line (false-color) and 3 mm continuum. The contour levels are $4\sigma_{\rm rms}$ ($1.6\,\mjbm$) to $84\sigma_{\rm rms}$ (peak) in step of $16\sigma_{\rm rms}$.  {\bf (b)} The $\htcop$ and $\chtoh$ emission regions around the main filament, overlaid on the IRAC-RGB image (3.6, 4.5, and 8.0 $\micron$ bands). The $\htcop$ contours are 15\% to 90\% in 15\%-step of the peak intensity (8.5 K $\kms$). The $\chtoh$ contours are 10\% to 90\% in 20\%-step of the peak intensity (18.7 K $\kms$). The dashed circles labels the area of each core. {\bf (c)} The spectra at the selected core centers. The blue and red-shaded areas indicate the velocity ranges of the two velocity components, respectively. The vertical dashed line denotes the division between the blue and main-filament components. }
    \label{fig:spec_img}    
\end{figure*}

\begin{figure*}[t]
    \centering
    \includegraphics[angle=0, width=1\textwidth]{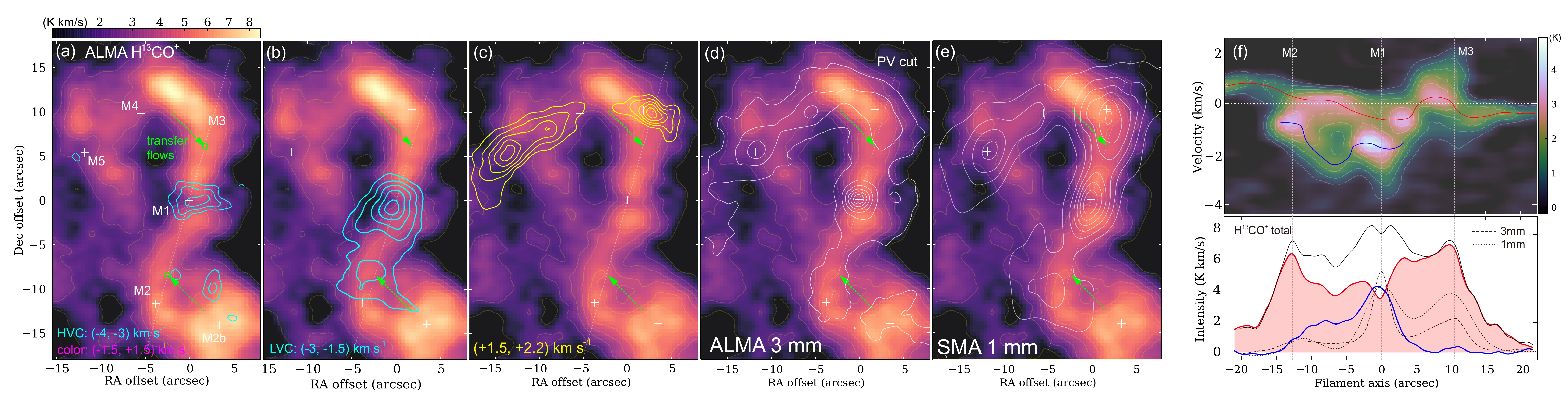} \\
    \caption{\small {\bf (a)} Contours: emission region of the $\htcop$ (1-0) line in three velocity intervals. The background image is the $\htcop$ emission in (-1.5,+1.5) $\kms$ (red-center component). For the blue-wing component, the contour levels are 4, 6, and 8 times of the rms level (0.3 K $\kms$). For the other components, the contour levels are 10\% to 90\% in 20\%-step of the peak intensity, which is 15, 11, and 3 K $\kms$ for the blue-center, red-center, and red-wing, respectively. The green arrows label the mass transfer flow directions onto the main filament (Chen21). The green circles labels the possible arrival points of the transfer flows onto the filament. {\bf (b)} PV plot and intensity profile along the major axis of the main filament. The sampling direction is labelled in dashed line in panel (a). The vertical dashed lines denote the projected offset of the three dense cores on the sampling direction. The horizontal dotted line represents the average systemic velocity of the filament. }
    \label{fig:m1_blue}    
\end{figure*}

\begin{figure*}[b]
    \centering
    \includegraphics[angle=0, width=0.6\textwidth]{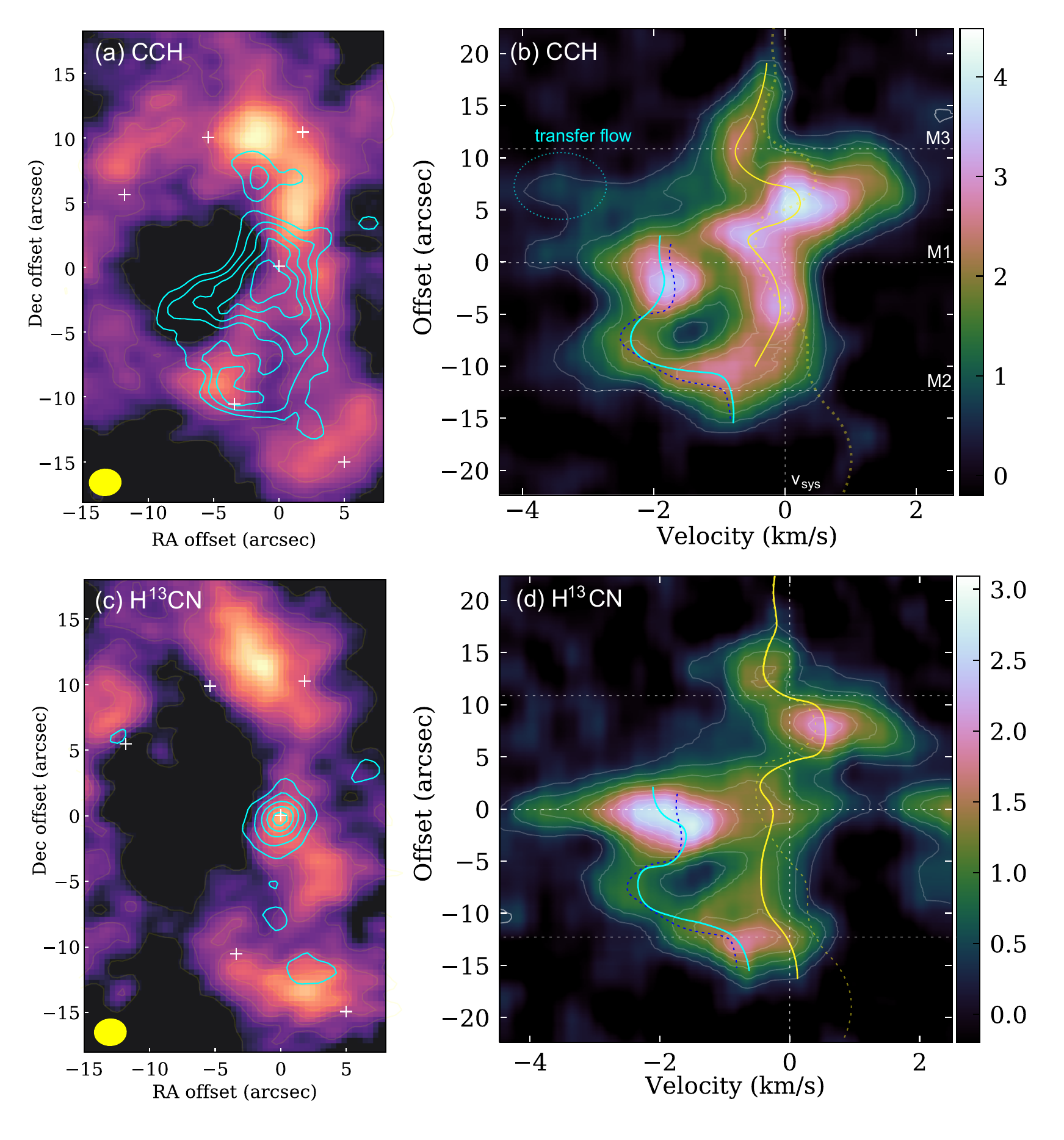} \\
    \caption{\small {\bf Left column:} The emission regions of the blue- (contours) and redshifted (false-color) components in CCS and $\htcn$ (1-0) lines. The contour levels are 20\% to 90\% of the peak intensity. {\bf Right column:} The PV diagrams of the two molecular lines. the $\htcn$ emission at $v_{\rm lsr}=2$ to 4 $\kms$ is from anther HFC of $F=1-1$. The HFCs of $\htcn$ are separated for 6-7 $\kms$ and would not blend with each other. The CCH (1-0) emission shows an additional small blueshift wings around M3, which should correspond to the transfer flow onto the filament (Chen21). This feature is not seen in other lines probably because of their lower optical depths. }
    \label{fig:pv_3mol}    
\end{figure*}

\begin{figure*}[b]
    \centering
    \includegraphics[angle=0, width=0.9\textwidth]{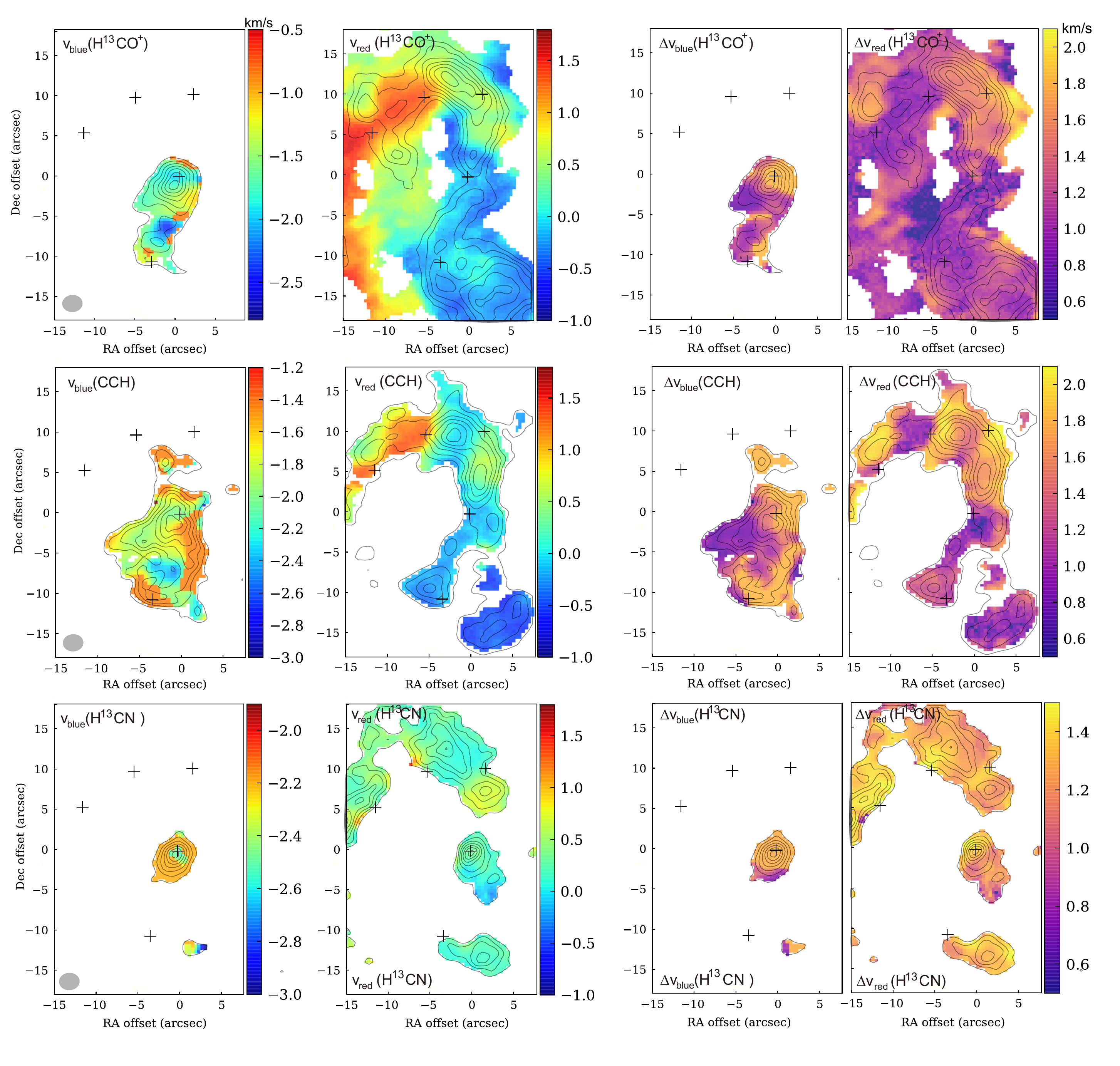} \\
    \caption{\small The peak velocity and line width distributions of the two components as observed in three molecular lines. The velocity ranges to measure the blueshift and main-filament components are $(-4, -1.8)$ and $(-1.5, 2)\ \kms$, respectively. {\bf left two columns:} The radial velocity distributions of the two components. {\bf right two columns:} The line-width distributions of the two components.}
    \label{fig:mom_3mol}    
\end{figure*}

\begin{figure*}[t]
    \centering
    \includegraphics[angle=0, width=0.8\textwidth]{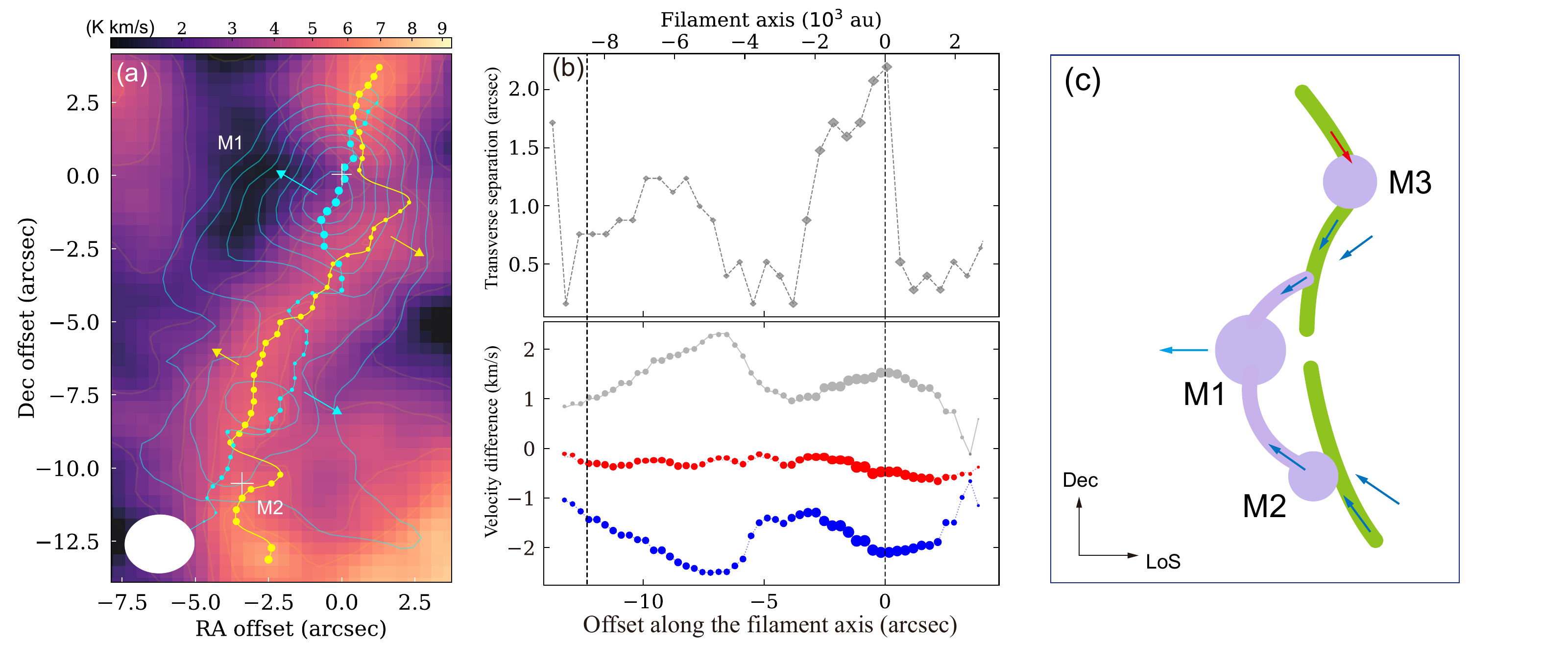} \\
    \caption{\small {\bf (a)} The ridge lines of the two velocity components, blue component in blue line and dots, red-center component (main filament) in yellow line and dots. The dot size is proportional to the integrated intensity. The arrows denote the projected moving direction of each component expected on the sky-plane. {\bf (b)} The transverse offset and velocity difference between the ridge lines of the two components along the filament axis. In the lower panel, the blue and red dots represent the peak radial velocity profile of the two components, respectively. The gray dots represent $\vert v_{\rm blue}-v_{\rm red} \vert$. {\bf (c)} A schematic view of the gas motion on the filament and cores. }
    \label{fig:ridge_dv}
\end{figure*}

\clearpage
\appendix
\section{Deriving the dense core properties} \label{app:parameters}
The total column density of a molecular species is estimated from the optically thin lines \citep{caselli02,henshaw14} as
\begin{eqnarray}
N_{\rm tot,mol}=\frac{8\pi I_{\rm tot}}{\lambda^{3}A}\frac{g_{u}}{g_{l}}\frac{1}{J_{\nu}(T_{ex})-J_{\nu}(T_{\rm bg})} \nonumber \\ \times\frac{1}{1-\exp{(-h\nu/k_{B}T_{\rm ex})}}\nonumber \\ \times \frac{Q_{\rm rot}(T_{\rm ex})}{g_{l}\exp{(-E{l}/k_{B}T_{\rm ex})}},
\end{eqnarray}
wherein $I_{\rm tot}\,=\,\int{T_{\rm b}dv}$ is the total intensity of the line profile, $A$ is the Einstein coefficient for spontaneous decay, $g_{u}$ and $g_{l}$ are the statistical weights for the upper and lower states, respectively. $J(T)=T_0/[\exp(T_0/T)-1]$ is the Planck-corrected brightness temperature with the reference value of $T_0=h\nu/k$. $Q_{rot}(T_{ex})$ is the partition function, $\lambda$ and $\nu$ are the wavelength and frequency of the line transition. $T_{\rm bg}=2.73$ K is the cosmic background temperature. $k_{B}$ and $h$ are the Boltzmann and Planck constants, respectively. The total gas column density is $N(\htwo)=N_{\rm tot,mol}/X_{\rm mol}$, wherein $X_{\rm mol}$ is the molecular abundance. For the $\htcop$ abundance, \citet{gerner14} measured $X_{\rm mol}(\htcop)=0.9$ to $1.5\times 10^{-9}$ from a large sample of dense cores from IRDC to hot-core stage. We adopt $X_{\rm mol}(\htcop)=(1.2\pm0.3)\times 10^{-9}$ for the calculation of G352. 

We also tried to use CCH and $\htcn$ (1-0) lines to calculate $N_{\rm tot}$. They provide comparable values on $10^{23}$ cm$^{-2}$ scale. The optical depth of the molecular lines are examined using \citep{Bell14}:
\begin{equation}
	\tau = \frac{c^3 A_{ij}}{8 \pi \nu^3}\frac{g_u N_{\rm tot}}{\Delta v Q_{\rm rot}} \exp\left(-\frac{E_{\rm up}}{k T_{\rm rot}}\right)\left[\exp\left(\frac{h \nu_{ij}}{k T_{\rm rot}}\right) - 1 \right].
\end{equation}
In calculation, we also adopted the average abundances in the young high-mass protostellar objects (HMPOs) in \citet{gerner14}, which are $X_{\rm mol}(\cch)=5.0\times 10^{-8}$ and $X_{\rm mol}(\htcn)=X_{\rm mol}(\hcn)/80=4.7\times 10^{-11}$. From the physical condition at M1 of $N_{\rm tot,mol}=2.6\times10^{23}\,\sqc$ and $T_{\rm ex}=37$ K, we estimated $\tau(\cch)=1.1$, $\tau(\htcop)=0.12$, and $\tau(\htcn)=0.06$. 

From $N(\htwo)$ distribution, we can estimate the core mass to be $m_{\rm core}=\mu m_{\rm H} \int N(\htwo) dA$, wherein the integration is made over the core area. The cores are irregular and not fully separated from the filament. To prevent complexity, we assumed the cores to have spherical or elliptical shapes. M3 and M5 indeed have noticeable elongation thus are assumed to be elliptical, while other cores are considered to be spherical. The core region is adjusted to cover the emission region above the extended emission over the filament (3 K $\kms$). The core areas are plotted in Figure \ref{fig:spec_img}a.

In deriving $N_{\rm tot,mol}$, we adopt the excitation temperature of $\hcop$ (1-0) lines (Figure \ref{fig:spec_img}). Assuming optically thick, it can be estimated using 
\begin{equation} \label{equ:jtex}
T_{\rm b}=J(T_{\rm ex})-J(T_{\rm bg}).	
\end{equation}
The calculation leads to $T_{\rm ex}\simeq\ 37$ K for M1 and $22\pm 3$ K for other cores. The second value is comparable to the dust temperature derived from SED fitting (Chen21), whereas the higher value at M1 should reflect the stellar heating. From $\chtcn$ population diagram, Chen21 derived a much higher temperature of $T_{\rm rot}=165$ K at M1. This value should incline to the small amount of shocked gas, whereas the major fraction of the filament should still be much cooler due to the absence of ionized gas and strong IR emission. Their temperature range is comparable with the values of pre-stellar cold dense cores \citep[e.g.][]{xie21}, suggesting an early evolutionary stage on average for the filament. Moreover, if assuming $T_{\rm ex}(\htcop)=165$ K, one can derive $N(\htwo)=7\times 10^{23}\,\sqc$ and $m_{\rm core}=30\,\msun$ for M1. This value is unreasonably high as it takes up more than 1/2 of the total filament mass. 

The total velocity dispersion of the core is estimated from the observed line width $\Delta v$ as 
\begin{equation} \label{equ:n_tot}
\begin{aligned}
\sigma_{\rm tot} & = \sqrt{\sigma_{\rm th}^2 + \sigma_{\rm nt}^2} \\
\quad            & = \sqrt{\frac{k_{\rm B} T_{\rm kin}}{\mu m_{\rm H}} + \frac{\Delta v^2}{8 \ln 2} - \frac{k_{\rm B} T_{\rm kin}}{m_{\rm mol}}},
\end{aligned}
\end{equation}
wherein $m_{\rm mol}$ is the molecular mass. The thermal part is 
\begin{equation}
\sigma_{\rm th}=\sqrt{\frac{k_{\rm B} T_{\rm kin}}{\mu m_{\rm H}}},
\end{equation}
and the non-thermal part is 
\begin{equation}
\sigma_{\rm nt} = \sqrt{\frac{\Delta v^2}{8 \ln 2} - \frac{k_{\rm B} T_{\rm kin}}{m_{\rm mol}}},
\end{equation}
In calculation we adopted a kinetic temperature of $T_{\rm kin}=37$ K for M1 and $T_{\rm kin}=22$ K for other cores. They lead to $\sigma_{\rm th}=0.29$ to 0.36 $\kms$, respectively. In comparison, the average line width of $\Delta v = 1.5\,\kms$ leads to $\sigma_{\rm nt}=0.64\,\kms$, which is not largely affected by $T_{\rm kin}$ and $\sigma_{\rm th}$.

\section{Outflow properties} \label{app:lines}
\begin{figure*}[b]
    \centering
    \includegraphics[angle=0, width=0.95\textwidth]{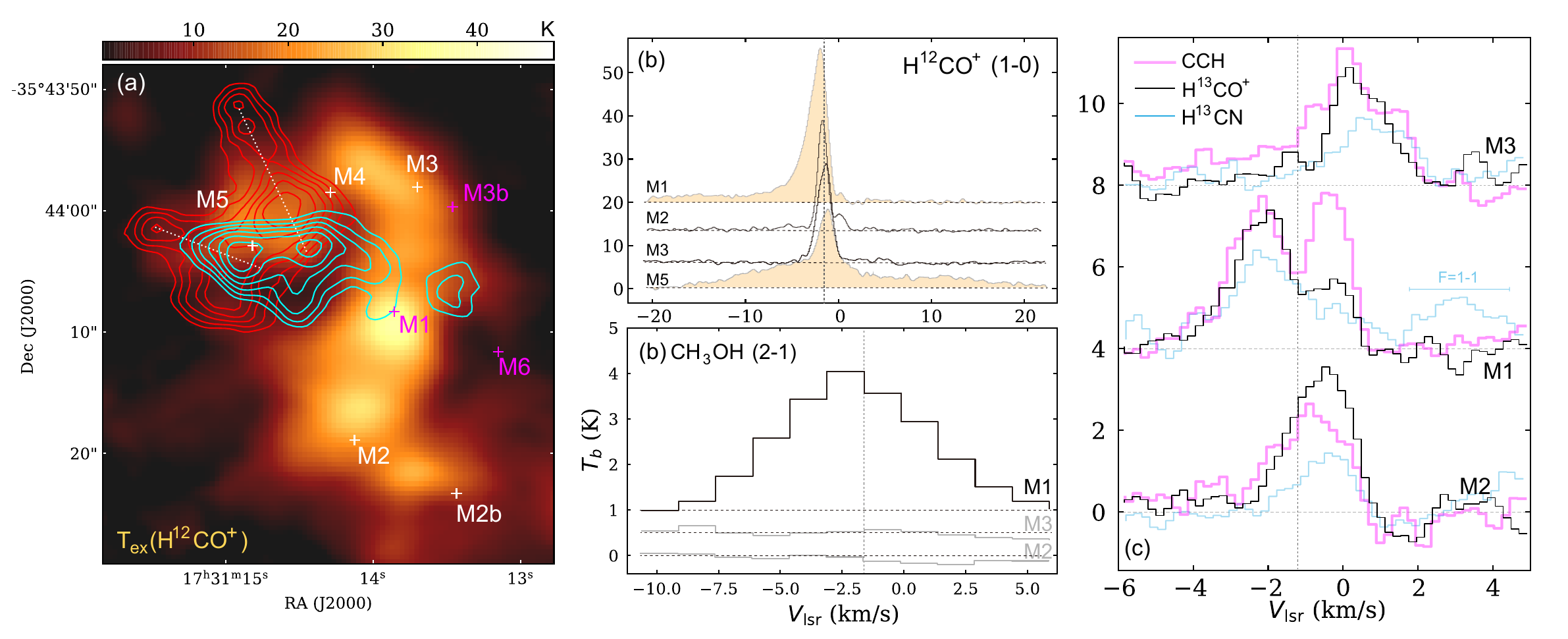} \\
    \caption{\small {\bf (a)} The $\hcop$ outflow lobes overlaid on the $\hcop$ excitation temperature map. The line intensity is converted to $T_{\rm ex}$ using Equation \ref{equ:jtex}. The contour levels are 15\% to 90\% of the maximum intensity, which is 22.6 and 22.9 K $\kms$ for the blue and red lobes, respectively. {\bf (b)(c)} The molecular spectra at selected cores. For each spectrum, the horizontal line represents the zero-level base line. The vertical dashed line denotes the division between the blue and main-filament components. }
    \label{fig:outflow_spec}    
\end{figure*}

The outflow emission is detected from high-velocity line wings in the $\hcop$ (1-0) lines. Its emission region and spectra are shown in Figure \ref{fig:outflow_spec}a and \ref{fig:outflow_spec}b, respectively. The blue- and redshift line wings are both strongly detected at M4, which shows the velocity ranges of $(-20,-6)$ $\kms$ for the blue lobe and $(5,20)$ $\kms$ for the red one. As seen in Figure \ref{fig:outflow_spec}a, compared with the CO outflow (Chen21), the $\hcop$ is concentrated around M4 and M5, suggesting that the outflow should be launched from this area. From the spatial extensions of the outflow lobes, we can see at least two bipolar outflows, as indicated by the dashed lines. The outflow extensions are not fully overlapped with M4 and M5. It is uncertain if this is due to additional driving sources or the time-variation of the outflow morphology. One thing for sure is that the outflow intensity is much weaker at M1, and the red lobe is almost undetected. If M1 also has a contribution to the blueshift gas, it should be more likely ejected in the process of core collapse and escape than due to the protostellar outflow. 

From the $\hcop$ line-wing intensities and outflow emission areas, they are found to have masses $m_{\rm blue}=0.20\,\msun$ and $m_{\rm red}=0.24\,\msun$. And the average radius is $R_{\rm out}=(8\pm 3)''=(5.5\pm 2)\times 10^3$ AU. Assuming an average velocity of $v_{\rm out}=8\,\kms$, the outflow age is derived to be $t_{\rm out}=R_{\rm out}/v_{\rm out}=(3.5\pm 1)\times 10^3$ years.    

\clearpage
\section{Acknowledgements}
The authors would like to thank the referees for the constructive comments. This work is supported by the National Key R\&D Program of China No.2022YFA1602901, National Natural Science Foundation of China (NSFC Grant Nos. 12041302, 11973051, U1931117, 11988101, 11725313, U2031117, and 12103069), and the International Partnership Program of the Chinese Academy of Sciences (grant No. 114A11KYSB
20210010). T.L. acknowledges the supports by the international partnership program of Chinese Academy of Sciences through grant No.114231KYSB20200009, and Shanghai Pujiang Program 20PJ1415500. L.B. gratefully acknowledges support by ANID BASAL project FB210003. A.S. gratefully acknowledges support by the Fondecyt Regular (project code 1220610), and ANID BASAL projects ACE210002 and FB210003. C.W. was supported by the Basic Science Research Program through the National Research Foundation of Korea (NRF) funded by the Ministry of Education, Science and Technology (NRF- 2019R1A2C1010851), and by the Korea Astronomy and Space Science Institute grant funded by the Korea government (MSIT; project No. 2023-1-84000). C.E  acknowledges the financial support from grant RJF/2020/000071 as a part of the Ramanujan Fellowship awarded by the Science and Engineering Research Board (SERB), India. E.M. is funded by the University of Helsinki doctoral school in particle physics and universe sciences (PAPU).

This paper makes use of the following ALMA data: ADS/JAO.ALMA 2019.1.00685.S. ALMA is a partnership of ESO (representing its member states), NSF (USA), and NINS (Japan), together with NRC (Canada), MOST and ASIAA (Taiwan region), and KASI (Republic of Korea), in cooperation with the Republic of Chile. The Joint ALMA Observatory is operated by ESO, AUI/NRAO, and NAOJ.

The Spitzer GLIMPSE legacy survey data can be downloaded at NASA/IPAC Infrared Science Archive (https://irsa.ipac.caltech.edu/irsaviewer/).

\clearpage

\begin{thebibliography}{}
\expandafter\ifx\csname natexlab\endcsname\relax\def\natexlab#1{#1}\fi
\providecommand{\url}[1]{\href{#1}{#1}}

\bibitem[{{{\'A}lvarez-Guti{\'e}rrez}
  {et~al.}(2021){{\'A}lvarez-Guti{\'e}rrez}, {Stutz}, {Law}, {Reissl},
  {Klessen}, {Leigh}, {Liu}, \& {Reeves}}]{alvarez21}
{{\'A}lvarez-Guti{\'e}rrez}, R.~H., {Stutz}, A.~M., {Law}, C.~Y., {et~al.}
  2021, \apj, 908, 86

\bibitem[{{Anathpindika} \& {Francesco}(2021)}]{anathpindika21}
{Anathpindika}, S.~V., \& {Francesco}, J.~D. 2021, \mnras, 502, 564

\bibitem[{{Arzoumanian} {et~al.}(2013){Arzoumanian}, {Andr{\'e}}, {Peretto}, \&
  {K{\"o}nyves}}]{arzoumanian13}
{Arzoumanian}, D., {Andr{\'e}}, P., {Peretto}, N., \& {K{\"o}nyves}, V. 2013,
  \aap, 553, A119

\bibitem[{{Bally} {et~al.}(2020){Bally}, {Ginsburg}, {Forbrich}, \&
  {Vargas-Gonz{\'a}lez}}]{bally20}
{Bally}, J., {Ginsburg}, A., {Forbrich}, J., \& {Vargas-Gonz{\'a}lez}, J. 2020,
  \apj, 889, 178

\bibitem[{{Bell} {et~al.}(2014){Bell}, {Cernicharo}, {Viti}, {Marcelino},
  {Palau}, {Esplugues}, \& {Tercero}}]{Bell14}
{Bell}, T.~A., {Cernicharo}, J., {Viti}, S., {et~al.} 2014, \aap, 564, A114

\bibitem[{{Benjamin} {et~al.}(2003){Benjamin}, {Churchwell}, {Babler}, {Bania},
  {Clemens}, {Cohen}, {Dickey}, {Indebetouw}, {Jackson}, {Kobulnicky},
  {Lazarian}, {Marston}, {Mathis}, {Meade}, {Seager}, {Stolovy}, {Watson},
  {Whitney}, {Wolff}, \& {Wolfire}}]{Benjamin03}
{Benjamin}, R.~A., {Churchwell}, E., {Babler}, B.~L., {et~al.} 2003, \pasp,
  115, 953

\bibitem[{{Bertoldi} \& {McKee}(1992)}]{bertoldi92}
{Bertoldi}, F., \& {McKee}, C.~F. 1992, \apj, 395, 140

\bibitem[{{Beuther} {et~al.}(2021){Beuther}, {Gieser}, {Suri}, {Linz},
  {Klaassen}, {Semenov}, {Winters}, {Henning}, {Soler}, {Urquhart}, {Syed},
  {Feng}, {M{\"o}ller}, {Beltr{\'a}n}, {S{\'a}nchez-Monge}, {Longmore},
  {Peters}, {Ballesteros-Paredes}, {Schilke}, {Moscadelli}, {Palau},
  {Cesaroni}, {Lumsden}, {Pudritz}, {Wyrowski}, {Kuiper}, \&
  {Ahmadi}}]{Beuther21}
{Beuther}, H., {Gieser}, C., {Suri}, S., {et~al.} 2021, \aap, 649, A113

\bibitem[{{Bhadari} {et~al.}(2020){Bhadari}, {Dewangan}, {Pirogov}, \&
  {Ojha}}]{bhadari20}
{Bhadari}, N.~K., {Dewangan}, L.~K., {Pirogov}, L.~E., \& {Ojha}, D.~K. 2020,
  \apj, 899, 167

\bibitem[{{Cantat-Gaudin} {et~al.}(2019){Cantat-Gaudin}, {Jordi}, {Wright},
  {Armstrong}, {Vallenari}, {Balaguer-N{\'u}{\~n}ez}, {Ramos}, {Bossini},
  {Padoan}, {Pelkonen}, {Mapelli}, \& {Jeffries}}]{gaudin19}
{Cantat-Gaudin}, T., {Jordi}, C., {Wright}, N.~J., {et~al.} 2019, \aap, 626,
  A17

\bibitem[{{Cao} {et~al.}(2022){Cao}, {Qiu}, {Zhang}, \& {Li}}]{cao22}
{Cao}, Y., {Qiu}, K., {Zhang}, Q., \& {Li}, G.-X. 2022, \apj, 927, 106

\bibitem[{{Caselli} {et~al.}(2002){Caselli}, {Walmsley}, {Zucconi}, {Tafalla},
  {Dore}, \& {Myers}}]{caselli02}
{Caselli}, P., {Walmsley}, C.~M., {Zucconi}, A., {et~al.} 2002, \apj, 565, 344

\bibitem[{{Chen} {et~al.}(2019){Chen}, {Zhang}, {Wright}, {Busquet}, {Lin},
  {Liu}, {Olguin}, {Sanhueza}, {Nakamura}, {Palau}, {Ohashi}, {Tatematsu}, \&
  {Liao}}]{chen19a}
{Chen}, H.-R.~V., {Zhang}, Q., {Wright}, M.~C.~H., {et~al.} 2019, \apj, 875, 24

\bibitem[{{Chen} {et~al.}(2021){Chen}, {Ren}, {Li}, {Liu}, {Wang}, {Shen},
  {Ellingsen}, {Sobolev}, {Mei}, {Li}, {Wu}, \& {Kim}}]{chen21}
{Chen}, X., {Ren}, Z.-Y., {Li}, D.-L., {et~al.} 2021, \apjl, 923, L20

\bibitem[{{Ching} {et~al.}(2018){Ching}, {Lai}, {Zhang}, {Girart}, {Qiu}, \&
  {Liu}}]{ching18}
{Ching}, T.-C., {Lai}, S.-P., {Zhang}, Q., {et~al.} 2018, \apj, 865, 110

\bibitem[{{Churchwell} {et~al.}(2009){Churchwell}, {Babler}, {Meade},
  {Whitney}, {Benjamin}, {Indebetouw}, {Cyganowski}, {Robitaille}, {Povich},
  {Watson}, \& {Bracker}}]{Churchwell09}
{Churchwell}, E., {Babler}, B.~L., {Meade}, M.~R., {et~al.} 2009, \pasp, 121,
  213

\bibitem[{{Clarke} \& {Whitworth}(2015)}]{Clarke15}
{Clarke}, S.~D., \& {Whitworth}, A.~P. 2015, \mnras, 449, 1819

\bibitem[{{Clarke} {et~al.}(2018){Clarke}, {Whitworth}, {Spowage},
  {Duarte-Cabral}, {Suri}, {Jaffa}, {Walch}, \& {Clark}}]{Clarke18}
{Clarke}, S.~D., {Whitworth}, A.~P., {Spowage}, R.~L., {et~al.} 2018, \mnras,
  479, 1722

\bibitem[{{Dharmawardena} {et~al.}(2022){Dharmawardena}, {Bailer-Jones},
  {Fouesneau}, {Foreman-Mackey}, {Coronica}, {Colnaghi}, {M{\"u}ller}, \&
  {Henshaw}}]{dharmawardena22}
{Dharmawardena}, T.~E., {Bailer-Jones}, C.~A.~L., {Fouesneau}, M., {et~al.}
  2022, arXiv e-prints, arXiv:2210.03615

\bibitem[{{Duan} {et~al.}(2023){Duan}, {Li}, {Goldsmith}, {Pagani}, {Ching},
  {Liu}, {Xie}, \& {Wang}}]{duan23}
{Duan}, Y., {Li}, D., {Goldsmith}, P.~F., {et~al.} 2023, \apj, 943, 182

\bibitem[{{Ducourant} {et~al.}(2017){Ducourant}, {Teixeira}, {Krone-Martins},
  {Bontemps}, {Despois}, {Galli}, {Bouy}, {Le Campion}, {Rapaport}, \&
  {Cuillandre}}]{ducourant17}
{Ducourant}, C., {Teixeira}, R., {Krone-Martins}, A., {et~al.} 2017, \aap, 597,
  A90

\bibitem[{{Fernandes} {et~al.}(2019){Fernandes}, {Montmerle}, {Santos-Silva},
  \& {Gregorio-Hetem}}]{fernandes19}
{Fernandes}, B., {Montmerle}, T., {Santos-Silva}, T., \& {Gregorio-Hetem}, J.
  2019, \aap, 628, A44

\bibitem[{{Galli} {et~al.}(2020){Galli}, {Bouy}, {Olivares}, {Miret-Roig},
  {Vieira}, {Sarro}, {Barrado}, {Berihuete}, {Bertout}, {Bertin}, \&
  {Cuillandre}}]{galli20}
{Galli}, P.~A.~B., {Bouy}, H., {Olivares}, J., {et~al.} 2020, \aap, 643, A148

\bibitem[{{Gerner} {et~al.}(2014){Gerner}, {Beuther}, {Semenov}, {Linz},
  {Vasyunina}, {Bihr}, {Shirley}, \& {Henning}}]{gerner14}
{Gerner}, T., {Beuther}, H., {Semenov}, D., {et~al.} 2014, \aap, 563, A97

\bibitem[{{Getman} {et~al.}(2019){Getman}, {Feigelson}, {Kuhn}, \&
  {Garmire}}]{getman19}
{Getman}, K.~V., {Feigelson}, E.~D., {Kuhn}, M.~A., \& {Garmire}, G.~P. 2019,
  \mnras, 487, 2977

\bibitem[{{Gro{\ss}schedl} {et~al.}(2018){Gro{\ss}schedl}, {Alves}, {Meingast},
  {Ackerl}, {Ascenso}, {Bouy}, {Burkert}, {Forbrich}, {F{\"u}rnkranz},
  {Goodman}, {Hacar}, {Herbst-Kiss}, {Lada}, {Larreina}, {Leschinski},
  {Lombardi}, {Moitinho}, {Mortimer}, \& {Zari}}]{grobschedl18}
{Gro{\ss}schedl}, J.~E., {Alves}, J., {Meingast}, S., {et~al.} 2018, \aap, 619,
  A106

\bibitem[{{Gupta} \& {Chen}(2022)}]{gupta22}
{Gupta}, A., \& {Chen}, W.-P. 2022, \aj, 163, 233

\bibitem[{{Ha} {et~al.}(2022){Ha}, {Li}, {Kounkel}, {Xu}, {Li}, \&
  {Zheng}}]{ha22}
{Ha}, T., {Li}, Y., {Kounkel}, M., {et~al.} 2022, \apj, 934, 7

\bibitem[{{Ha} {et~al.}(2021){Ha}, {Li}, {Xu}, {Kounkel}, \& {Li}}]{ha21}
{Ha}, T., {Li}, Y., {Xu}, S., {Kounkel}, M., \& {Li}, H. 2021, \apjl, 907, L40

\bibitem[{{Hacar} {et~al.}(2017){Hacar}, {Tafalla}, \& {Alves}}]{hacar17}
{Hacar}, A., {Tafalla}, M., \& {Alves}, J. 2017, \aap, 606, A123

\bibitem[{{Henshaw} {et~al.}(2014){Henshaw}, {Caselli}, {Fontani},
  {Jim{\'e}nez-Serra}, \& {Tan}}]{henshaw14}
{Henshaw}, J.~D., {Caselli}, P., {Fontani}, F., {Jim{\'e}nez-Serra}, I., \&
  {Tan}, J.~C. 2014, \mnras, 440, 2860

\bibitem[{{Herczeg} {et~al.}(2019){Herczeg}, {Kuhn}, {Zhou}, {Hatchell},
  {Manara}, {Johnstone}, {Dunham}, {Bhardwaj}, {Jose}, \& {Yuan}}]{herczeg19}
{Herczeg}, G.~J., {Kuhn}, M.~A., {Zhou}, X., {et~al.} 2019, \apj, 878, 111

\bibitem[{{Hu} {et~al.}(2021){Hu}, {Qiu}, {Cao}, {Liu}, {Wang}, {Li}, {Shen},
  {Li}, {Wang}, {Li}, \& {Dong}}]{hu21}
{Hu}, B., {Qiu}, K., {Cao}, Y., {et~al.} 2021, \apj, 908, 70

\bibitem[{{Jerabkova} {et~al.}(2019){Jerabkova}, {Boffin}, {Beccari}, \&
  {Anderson}}]{jerabkova19}
{Jerabkova}, T., {Boffin}, H. M.~J., {Beccari}, G., \& {Anderson}, R.~I. 2019,
  \mnras, 489, 4418

\bibitem[{{Kim} {et~al.}(2019){Kim}, {Lu}, {Konopacky}, {Chu}, {Toller},
  {Anderson}, {Theissen}, \& {Morris}}]{kim19}
{Kim}, D., {Lu}, J.~R., {Konopacky}, Q., {et~al.} 2019, \aj, 157, 109

\bibitem[{{Kounkel} {et~al.}(2022){Kounkel}, {Stassun}, {Covey}, \&
  {Hartmann}}]{kounkel22}
{Kounkel}, M., {Stassun}, K.~G., {Covey}, K., \& {Hartmann}, L. 2022, \mnras,
  517, 161

\bibitem[{{Kraus} {et~al.}(2017){Kraus}, {Herczeg}, {Rizzuto}, {Mann},
  {Slesnick}, {Carpenter}, {Hillenbrand}, \& {Mamajek}}]{kraus17}
{Kraus}, A.~L., {Herczeg}, G.~J., {Rizzuto}, A.~C., {et~al.} 2017, \apj, 838,
  150

\bibitem[{{Krolikowski} {et~al.}(2021){Krolikowski}, {Kraus}, \&
  {Rizzuto}}]{krolikowski21}
{Krolikowski}, D.~M., {Kraus}, A.~L., \& {Rizzuto}, A.~C. 2021, \aj, 162, 110

\bibitem[{{Kuhn} {et~al.}(2020){Kuhn}, {Hillenbrand}, {Carpenter}, \& {Avelar
  Menendez}}]{kuhn20}
{Kuhn}, M.~A., {Hillenbrand}, L.~A., {Carpenter}, J.~M., \& {Avelar Menendez},
  A.~R. 2020, \apj, 899, 128

\bibitem[{{Li} {et~al.}(2013){Li}, {Kauffmann}, {Zhang}, \& {Chen}}]{li13}
{Li}, D., {Kauffmann}, J., {Zhang}, Q., \& {Chen}, W. 2013, \apjl, 768, L5

\bibitem[{{Li} {et~al.}(2022){Li}, {Sanhueza}, {Lee}, {Zhang}, {Beuther},
  {Palau}, {Liu}, {Smith}, {Liu}, {Jim{\'e}nez-Serra}, {Kim}, {Feng}, {Liu},
  {Wang}, {Li}, {Qiu}, {Lu}, {Girart}, {Wang}, {Li}, {Li}, {Cao}, {Kim}, \&
  {Strom}}]{li22a}
{Li}, S., {Sanhueza}, P., {Lee}, C.~W., {et~al.} 2022, \apj, 926, 165

\bibitem[{{Liu} {et~al.}(2022){Liu}, {Tej}, {Liu}, {Goldsmith}, {Stutz},
  {Juvela}, {Qin}, {Xu}, {Bronfman}, {Evans}, {Saha}, {Issac}, {Tatematsu},
  {Wang}, {Li}, {Zhang}, {Baug}, {Dewangan}, {Wu}, {Zhang}, {Lee}, {Liu},
  {Zhou}, \& {Soam}}]{liu22}
{Liu}, H.-L., {Tej}, A., {Liu}, T., {et~al.} 2022, \mnras, 511, 4480

\bibitem[{{Liu} {et~al.}(2018){Liu}, {Li}, {Juvela}, {Kim}, {Evans}, {Di
  Francesco}, {Liu}, {Yuan}, {Tatematsu}, {Zhang}, {Ward-Thompson}, {Fuller},
  {Goldsmith}, {Koch}, {Sanhueza}, {Ristorcelli}, {Kang}, {Chen}, {Hirano},
  {Wu}, {Sokolov}, {Lee}, {White}, {Wang}, {Eden}, {Li}, {Thompson}, {Pattle},
  {Soam}, {Nasedkin}, {Kim}, {Kim}, {Lai}, {Park}, {Qiu}, {Zhang}, {Alina},
  {Eswaraiah}, {Falgarone}, {Fich}, {Greaves}, {Gu}, {Kwon}, {Li}, {Malinen},
  {Montier}, {Parsons}, {Qin}, {Rawlings}, {Ren}, {Tang}, {Tang}, {Toth},
  {Wang}, {Wouterloot}, {Yi}, \& {Zhang}}]{liu18b}
{Liu}, T., {Li}, P.~S., {Juvela}, M., {et~al.} 2018, \apj, 859, 151

\bibitem[{{Liu} {et~al.}(2020){Liu}, {Evans}, {Kim}, {Goldsmith}, {Liu},
  {Zhang}, {Tatematsu}, {Wang}, {Juvela}, {Bronfman}, {Cunningham}, {Garay},
  {Hirota}, {Lee}, {Kang}, {Li}, {Li}, {Mardones}, {Qin}, {Ristorcelli}, {Tej},
  {Toth}, {Wu}, {Wu}, {Yi}, {Yun}, {Liu}, {Peng}, {Li}, {Li}, {Lee}, {Shen},
  {Baug}, {Wang}, {Zhang}, {Issac}, {Zhu}, {Luo}, {Soam}, {Liu}, {Xu}, {Wang},
  {Zhang}, {Ren}, \& {Zhang}}]{liu20}
{Liu}, T., {Evans}, N.~J., {Kim}, K.-T., {et~al.} 2020, \mnras, 496, 2790

\bibitem[{{McKee} \& {Ostriker}(2007)}]{mckee07}
{McKee}, C.~F., \& {Ostriker}, E.~C. 2007, \araa, 45, 565

\bibitem[{{Megeath} {et~al.}(2012){Megeath}, {Gutermuth}, {Muzerolle},
  {Kryukova}, {Flaherty}, {Hora}, {Allen}, {Hartmann}, {Myers}, {Pipher},
  {Stauffer}, {Young}, \& {Fazio}}]{megeath12}
{Megeath}, S.~T., {Gutermuth}, R., {Muzerolle}, J., {et~al.} 2012, \aj, 144,
  192

\bibitem[{{Montillaud} {et~al.}(2019){Montillaud}, {Juvela}, {Vastel}, {He},
  {Liu}, {Ristorcelli}, {Eden}, {Kang}, {Kim}, {Koch}, {Lee}, {Rawlings},
  {Saajasto}, {Sanhueza}, {Soam}, {Zahorecz}, {Alina}, {B{\"o}gner}, {Cornu},
  {Doi}, {Malinen}, {Marshall}, {Micelotta}, {Pelkonen}, {Viktor T{\'o}th},
  {Traficante}, \& {Wang}}]{Montillaud19}
{Montillaud}, J., {Juvela}, M., {Vastel}, C., {et~al.} 2019, \aap, 631, A3

\bibitem[{{Ostriker}(1964)}]{ostriker64}
{Ostriker}, J. 1964, \apj, 140, 1056

\bibitem[{{Peretto} {et~al.}(2013){Peretto}, {Fuller}, {Duarte-Cabral},
  {Avison}, {Hennebelle}, {Pineda}, {Andr{\'e}}, {Bontemps}, {Motte},
  {Schneider}, \& {Molinari}}]{peretto13}
{Peretto}, N., {Fuller}, G.~A., {Duarte-Cabral}, A., {et~al.} 2013, \aap, 555,
  A112

\bibitem[{{Punanova} {et~al.}(2018){Punanova}, {Caselli}, {Pineda}, {Pon},
  {Tafalla}, {Hacar}, \& {Bizzocchi}}]{punanova18}
{Punanova}, A., {Caselli}, P., {Pineda}, J.~E., {et~al.} 2018, \aap, 617, A27

\bibitem[{{Quintana} \& {Wright}(2022)}]{quintana22}
{Quintana}, A.~L., \& {Wright}, N.~J. 2022, \mnras, 515, 687

\bibitem[{{Ren} {et~al.}(2021){Ren}, {Zhu}, {Shi}, {Yue}, {Li}, {Zhang},
  {Mardones}, {Wu}, {Jiao}, {Liu}, {Luo}, {Xie}, {Zhang}, \& {Xu}}]{ren21}
{Ren}, Z., {Zhu}, L., {Shi}, H., {et~al.} 2021, \mnras, 505, 5183

\bibitem[{{Rivera-Ortiz} {et~al.}(2021){Rivera-Ortiz},
  {Rodr{\'\i}guez-Gonz{\'a}lez}, {Cant{\'o}}, \& {Zapata}}]{ortiz21}
{Rivera-Ortiz}, P.~R., {Rodr{\'\i}guez-Gonz{\'a}lez}, A., {Cant{\'o}}, J., \&
  {Zapata}, L.~A. 2021, \apj, 916, 56

\bibitem[{{Roccatagliata} {et~al.}(2020){Roccatagliata}, {Franciosini},
  {Sacco}, {Randich}, \& {Sicilia-Aguilar}}]{roccatagliata20}
{Roccatagliata}, V., {Franciosini}, E., {Sacco}, G.~G., {Randich}, S., \&
  {Sicilia-Aguilar}, A. 2020, \aap, 638, A85

\bibitem[{{Sharma} {et~al.}(2020){Sharma}, {Gopinathan}, {Soam}, {Lee}, {Kim},
  {Ghosh}, {Tej}, {Kim}, {Sharma}, \& {Saha}}]{sharma20}
{Sharma}, E., {Gopinathan}, M., {Soam}, A., {et~al.} 2020, \aap, 639, A133

\bibitem[{{Shimajiri} {et~al.}(2019){Shimajiri}, {Andr{\'e}}, {Ntormousi},
  {Men'shchikov}, {Arzoumanian}, \& {Palmeirim}}]{shimajiri19}
{Shimajiri}, Y., {Andr{\'e}}, P., {Ntormousi}, E., {et~al.} 2019, \aap, 632,
  A83

\bibitem[{{Stutz} \& {Gould}(2016)}]{stutz16}
{Stutz}, A.~M., \& {Gould}, A. 2016, \aap, 590, A2

\bibitem[{{Swiggum} {et~al.}(2021){Swiggum}, {D'Onghia}, {Alves},
  {Gro{\ss}schedl}, {Foley}, {Zucker}, {Meingast}, {Chen}, \&
  {Goodman}}]{swiggum21}
{Swiggum}, C., {D'Onghia}, E., {Alves}, J., {et~al.} 2021, \apj, 917, 21

\bibitem[{{Szegedi-Elek} {et~al.}(2019){Szegedi-Elek}, {Kun}, {Mo{\'o}r},
  {Marton}, \& {Reipurth}}]{szegedi19}
{Szegedi-Elek}, E., {Kun}, M., {Mo{\'o}r}, A., {Marton}, G., \& {Reipurth}, B.
  2019, \mnras, 484, 1800

\bibitem[{{Tobin} {et~al.}(2009){Tobin}, {Hartmann}, {Furesz}, {Mateo}, \&
  {Megeath}}]{tobin09}
{Tobin}, J.~J., {Hartmann}, L., {Furesz}, G., {Mateo}, M., \& {Megeath}, S.~T.
  2009, \apj, 697, 1103

\bibitem[{{Tu} {et~al.}(2022){Tu}, {Zucker}, {Speagle}, {Beane}, {Goodman},
  {Alves}, {Faherty}, \& {Burkert}}]{tu22}
{Tu}, A.~J., {Zucker}, C., {Speagle}, J.~S., {et~al.} 2022, \apj, 936, 57

\bibitem[{{Walsh} {et~al.}(1998){Walsh}, {Burton}, {Hyland}, \&
  {Robinson}}]{walsh98}
{Walsh}, A.~J., {Burton}, M.~G., {Hyland}, A.~R., \& {Robinson}, G. 1998,
  \mnras, 301, 640

\bibitem[{{Ward} {et~al.}(2020){Ward}, {Kruijssen}, \& {Rix}}]{ward20}
{Ward}, J.~L., {Kruijssen}, J.~M.~D., \& {Rix}, H.-W. 2020, \mnras, 495, 663

\bibitem[{{Xie} {et~al.}(2021){Xie}, {Fuller}, {Li}, {Chen}, {Ren}, {Wu},
  {Duan}, {Wang}, {Li}, {Peretto}, {Liu}, \& {Shen}}]{xie21}
{Xie}, J., {Fuller}, G.~A., {Li}, D., {et~al.} 2021, arXiv e-prints,
  arXiv:2103.12985

\bibitem[{{Zamora-Avil{\'e}s} {et~al.}(2019){Zamora-Avil{\'e}s},
  {Ballesteros-Paredes}, {Hern{\'a}ndez}, {Rom{\'a}n-Z{\'u}{\~n}iga}, {Lora},
  \& {Kounkel}}]{zamora19}
{Zamora-Avil{\'e}s}, M., {Ballesteros-Paredes}, J., {Hern{\'a}ndez}, J.,
  {et~al.} 2019, \mnras, 488, 3406

\end{thebibliography}

\end{document}